\documentclass[conference]{IEEEtran}
\IEEEoverridecommandlockouts
\newcommand{\blue}[1]{{\textcolor[rgb]{0,0,1}{#1}}}

\usepackage{subfigure}
\usepackage{cite}
\usepackage{amsmath,amssymb,amsfonts}
\usepackage{algorithm}
\usepackage{algorithmic}
\usepackage{graphicx}
\usepackage{textcomp}
\usepackage{xcolor}
\usepackage{bm}
\usepackage{soul}
\def\BibTeX{{\rm B\kern-.05em{\sc i\kern-.025em b}\kern-.08em
    T\kern-.1667em\lower.7ex\hbox{E}\kern-.125emX}}

\IEEEaftertitletext{\vspace{-0.5\baselineskip}}
\begin{document}

\title {Efficient Near Maximum-Likelihood Reliability-Based Decoding for Short LDPC Codes\\}


\author{\IEEEauthorblockN{Weiyang Zhang, Chentao Yue, Yonghui Li, and Branka Vucetic}
\IEEEauthorblockA{School of Electrical and Information Engineering, The University of Sydney, Australia\\
\{weiyang.zhang, chentao.yue, yonghui.li, branka.vucetic\}@sydney.edu.au
}

}

\maketitle

\begin{abstract}
In this paper, we propose an efficient decoding algorithm for short low-density parity check (LDPC) codes by carefully combining the belief propagation (BP) decoding and order statistic decoding (OSD) algorithms. Specifically, a modified BP (mBP) algorithm is applied for a certain number of iterations prior to OSD to enhance the reliability of the received message, where an offset parameter is utilized in mBP to control the weight of the extrinsic information in message passing. By carefully selecting the offset parameter and the number of mBP iterations, the number of errors in the most reliable positions (MRPs) in OSD can be reduced by mBP, thereby significantly improving the overall decoding performance of error rate and complexity. Simulation results show that the proposed algorithm can approach the maximum-likelihood decoding (MLD) for short LDPC codes with only a slight increase in complexity compared to BP and a significant decrease compared to OSD. Specifically, the order-$(m-1)$ decoding of the proposed algorithm can achieve the performance of the order-$m$ OSD.
\end{abstract}

\vspace{-0.3em}
\section{Introduction}
\vspace{-0.3em}

Low density parity check (LDPC) codes have been widely applied in wireless communications \cite{b1}. In 1963, Gallager investigated the iterative belief propagation (BP) decoding algorithm for LDPC codes \cite{b2}. BP can achieve near-optimal performances for long LDPC codes with approximately cycle-free Tanner graphs \cite{b3,b4}, which has been verified in \cite{b5} by the density evolution analysis. However, BP is incapable of achieving maximum likelihood decoding (MLD) for short LDPC codes, due to the short cycles in their Tanner graphs. There have been many works on improving the block error rate (BLER) performance of BP when cycles present in Tanner graphs, including modifying the message-passing function \cite{b7,b7,b8,b9,b10}, and dynamic scheduling of message-passing strategies \cite{b11}. Despite these efforts, when dealing with short LDPC codes, BP still exhibits a significant error-rate performance gap compared to MLD.

As one of 5G key user scenarios, ultra-reliable and low latency communications (uRLLC) requires short codes with low decoding complexity and superior error rate performance, to satisfy its stringent requirements of reliability and latency\cite{b12,b13}. As reviewed in \cite{b14}, candidate code/decoder pairs for uRLLC include short polar codes with successive cancellation list (SCL) decoding and short Bose-Chaudhuri-Hocquenghem (BCH) codes with ordered statistics decoding (OSD). Until recently, short LDPC was not considered suitable for uRLLC due to the inferior BLER performance of BP decoding, despite its very low decoding complexity. For example, the half-rate length-128 LDPC code shows a 1.75dB gap to the normal approximation (NA) bound \cite{b15} under BP. However, this gap can be shortened to 0.5 dB under MLD \cite{b16}. This implies that short LDPC codes will also be a competitive candidate for uRLLC if a low-complexity near-MLD decoder is available.

Combining BP and OSD algorithms is a potential solution to the inferior BLER performance of the BP decoding. OSD was firstly proposed in 1995 as an approximate MLD algorithm \cite{b17}. An OSD with order $m$ has complexity as high as $O(K^m)$ for $K$ information bits. Although OSD has been significantly improved in recent years \cite{b18,b19}, its complexity remains high at high SNRs when compared to SCL and BP \cite{b14}. In \cite{b20}, Fossorier proposed to iteratively perform BP and OSD for medium-length LDPC codes to achieve the near-MLD performance. However, this algorithm requires multiple Gaussian elimination operations in a single decoding round, leading to high decoding latency. A decoding algorithm concatenating BP decoding and OSD was proposed in \cite{b21}. This algorithm uses the log-likelihood ratio (LLR) accumulated from each BP iteration as the input of OSD. Nevertheless, it necessitates a high decoding order of OSD to achieve MLD for short codes, because short cycles can introduce message correlations, thereby undermining the reliability of the OSD input. In \cite{b22}, a simple concatenation of BP and OSD was also used for decoding non-binary short LDPC codes. The complexity of these MLD methods remains high for uRLLC.

This paper proposes a novel modified-BP-OSD (mBP-OSD) decoding algorithm for short LDPC codes. Firstly, the original BP decoding is performed to decode the received sequence. If the output of BP satisfies a stopping criterion, it will be directly output as the final decoding result. Otherwise, we perform the modified BP (mBP) with the offset parameter $\beta$ over the received sequence for $\alpha$ iterations, where $\alpha$ is determined by the girth of code Tanner graph and $\beta$ controls the weight of extrinsic information during message passing. Then, based on the LLR output by mBP, an order-$m$ OSD is conducted to obtain the final result. Simulations show that the proposed mBP-OSD can achieve near-MLD performance for short LDPC codes, while having significantly lower complexity than OSD and only a slightly higher complexity than BP. Furthermore, we highlight that mBP-OSD can achieve the error rate of order-$m$ OSD approaches with the order-$(m-1)$ decoding procedure for short LDPC codes. Thus, any complexity-reduced OSD approaches, such as \cite{b18,b23}, can be applied in mBP-OSD to further reduce their complexity.

The rest of this paper is organized as follows: Section II describes preliminaries. Section III elaborates on the proposed mBP-OSD algorithm. Simulation results are presented in Section IV and conclusions are provided in Section V.
\vspace{-0.3em}
\section{Preliminaries}
\vspace{-0.3em}
We consider a binary linear block code $\mathcal{C}(N, K)$ defined by the generator matrix $\mathbf{G}$ and parity check matrix $\mathbf{H}$, where $N$ and $K$ denote the codeword length and the information block size, respectively. A codeword $\mathbf{c} = [c_1,\ldots,c_N]$ of $\mathcal{C}(N, K)$ is transmitted over the additive white Gaussian noise (AWGN) channel with binary phase shift keying (BPSK) modulation. Let $\mathbf{w} = [w_1,
\ldots,w_N]$ denote the noise vector with each element having zero mean and variance of $N_0/2$, where $N_0$ is the single-band noise power spectrum density. Thus, the received signal is $\mathbf{r}=\mathbf{s}+\mathbf{w}$, where $\mathbf{s} = 1 - 2\mathbf{c}$ is the BPSK symbol vector. The signal-to-noise ratio (SNR) is accordingly defined as $\rm{SNR}= 2/\emph{N}_0$.

\vspace{-0.3em}
\subsection{Belief Propagation Decoding}
\vspace{-0.3em}

The BP decoding is accomplished through the information exchange between $N$ variable nodes $v_i$ for $i\in\{1,\ldots,N\}$ and $N-K$ check nodes $c_j$ for $j\in\{{1,\ldots,N-K\}}$ in the Tanner graph. We refer readers to a detailed introduction of BP in \cite{b24}. Let us denote the channel LLR of variable nodes as $\ell=[\ell_1,\ell_2,\ldots,\ell_N]$, which is give by $\ell_i=4r_i/N_0$ for BPSK symbols. The absolute value of LLR, i.e., $|\ell_i|$, represents the reliability of the $i$-th received bit. During the decoding, BP iteratively passes messages between variable nodes and check nodes along the edges of the Tanner graph. Let $\mathcal{A}_i$ denote the set of check notes connected with $i$-th variable node in Tanner graph and $\mathcal{B}_j$ denote the set of variable nodes connected with $j$-th check node. The message passing procedure in each iteration is described as
\begin{equation}
M_{ji}=\ell_i+\sum_{j^{'}\in{\mathcal{A}_i}\backslash j}E_{ij^{'}} \blue{,}
\end{equation}
for all $i\in\{1,\ldots,N\}$ and $j\in \mathcal{A}_i$, and
\begin{equation}
E_{ij}=2\tanh^{-1}(\prod_{i^{'}\in{\mathcal{B}_j}\backslash i}\tanh(M_{ji^{'}}/2)),
\end{equation}
for all $j\in\{1,\ldots,N-K\}$ and $i\in \mathcal{B}_i$. In (1) and (2), $M_{ji}$ denotes the messages sent from the variable nodes $v_i$ to the check nodes $c_j$, and $E_{ij}$ denotes the messages returned from the check node $c_j$ to the variable node $v_i$. Particularly, for the first iteration, $E_{ij}=0$ for all $i$ and $j$. After each iteration, BP computes the posterior LLR $\mathbf{L}=[L_1,L_2,\ldots,L_N]$ of the variable nodes according to

\begin{equation}
L_i=\ell_i+\sum_{j\in{\mathcal{A}_i}}E_{ij}.
\end{equation}

Then, the hard decision is performed over $\mathbf{L}$ to obtain a binary codeword estimate $\mathbf{x}$, according to $x_i = 0$ if $L_i\geq 0$, and $x_i = 1$, otherwise. The BP iteration will be terminated when $\mathbf{x}\mathbf{H}^{\rm{T}}=0$ or a maximum number of iterations $\rm{T}_{\max}$ is reached. Upon termination, $\mathbf{x}$ is output as the final decoding result. In general, for each bit (variable node), the channel LLR, i.e., $\ell_{i}$, is referred to as the internal information, while the message passed from check nodes, i.e., $E_{ij}$ in (1), is referred to as the extrinsic information.

\vspace{-0.3em}
\subsection{Ordered Statistic Decoding}
 \vspace{-0.3em}

With minor notation abuse, let $\bm{\ell} = [\ell_1\ldots,\ell_N]$ denote the input LLR to OSD. In the first step of OSD, a permutation ${\pi_1}$ is applied to $\bm{\ell}$ and the columns of the generator matrix $\textbf{G}$ to sort them as $\bm{\ell}^{'}=\pi_1(\bm{\ell})$ and $\textbf{G}^{'}=\pi_1(\textbf{G})$, respectively. After the permutation $\pi_1$, the sorted LLR $\bm{\ell}'$ should satisfy
\begin{equation}
\lvert{\ell^{'}_1}\rvert\geq\lvert{\ell^{'}_2}\rvert\geq\ldots\geq\lvert{\ell^{'}_{N-1}}\rvert\geq\lvert{\ell^{'}_{N}}\rvert .
\end{equation}

Next, Gaussian elimination is performed on the matrix $\mathbf{G}^{'}$ to obtain the systematic form $\mathbf{G}^{''}$. During this process, an additional permutation ${\pi_2}$ is applied to ensure the first $K$ columns of the matrix are linearly independent. The same permutation is applied to $\bm{\ell}'$ to obtain $\bm{\ell}^{''}={\pi_2}({\pi_1}(\bm{\ell}))$ which satisfies the inequalities:

\begin{equation}
\lvert{\ell^{''}_1}\rvert\geq\lvert{\ell^{''}_2}\rvert\geq\ldots\geq\lvert{\ell^{''}_K}\rvert
\ \ \text{and} \ \ 
\lvert{\ell^{''}_{K+1}}\rvert\geq\lvert{\ell^{''}_{K+2}}\rvert\geq\ldots\geq\lvert{\ell^{''}_{N}}\rvert.
\end{equation}

After the permutations, the first $K$ positions of $\bm{\ell}''$ is defined as the most reliable positions (MRPs). Let $\mathbf{y}=[y_1,y_2,...,y_N]$ denote the hard decision of $\bm{\ell}''$ and $\mathbf{y}_\mathrm{B}=[y_1,\ldots,y_K]$ denote the first $K$ positions of $\mathbf{y}$ corresponding to MRPs. Then, OSD performs ``reprocessing'' to generate codeword candidates. An order-$m$ OSD algorithm consists $m+1$ reprocessing phases from phase-0 to phase-$m$. For the phase-0 processing, a codeword candidate is obtained as
\begin{equation}
\mathbf{c}=\mathbf{y}_\mathrm{B}\mathbf{G}^{''}.
\end{equation}
We note that $\mathbf{c}$ is the correct codeword if and only if there are no errors in MRPs \cite{b17}. For the phase-$q$ processing for $1\leq q \leq m$, all binary test error patterns (TEP) $\mathbf{e}=[e_1,e_2,...,e_K]$ containing $q$ non-zero positions are applied to modify $\mathbf{y}_{\mathrm{B}}$ according to
\begin{equation}
\mathbf{y}_{\mathrm{e}}=\mathbf{y}_\mathrm{B}{\oplus}\mathbf{e}.
\end{equation}
Then, similar to (6), a new candidate codeword $\mathbf{c}_\mathrm{e}$ is obtained as $\mathbf{c}_{\mathrm{e}} = \mathbf{y}_{\mathrm{e}}\mathbf{G}^{''}$ for each TEP $\mathbf{e}$ in the phase-$q$ reprocessing. 
After all the $m+1$ reprocessing phases, the total number of candidate codewords is $\sum_{i=0}^{m}\binom{K}{i}$. OSD finds the best codeword estimate $\mathbf{c}_{\min}$ with the minimum weighted Hamming distance (WHD) among all these candidate codewords. The WHD from a codeword $\mathbf{c}_{\mathrm{e}}$ to $\mathbf{y}$ is defined as
\begin{equation}
\mathcal{D}_{\mathrm{e}}=\sum_{\substack{0<i<N \\c_{\mathrm{e},i}\ne{y_i}}}|\ell_i^{''}|.
\end{equation}
Finally, the permutations $\pi_1$ and $\pi_2$ are restored to get the decoding result, i.e., $\textbf{c}_{\mathrm{opt}}=\pi_1^{-1}(\pi_2^{-1}(\textbf{c}_{\min}))$.

As reported in \cite{b2}, higher decoding order indicates a better BLER performance, and for a code with the minimum Hamming distance $d_{\mathrm{H}}$, OSD with order $m=\lfloor{d_{\mathrm{H}}/4}\rfloor$ can approximately achieve MLD.

\vspace{-0.3em}
\section{Modified BP-OSD Algorithm}
\vspace{-0.3em}

\subsection{Algorithm Overview}
\begin{figure}
\centering
\includegraphics[width=0.85\linewidth]{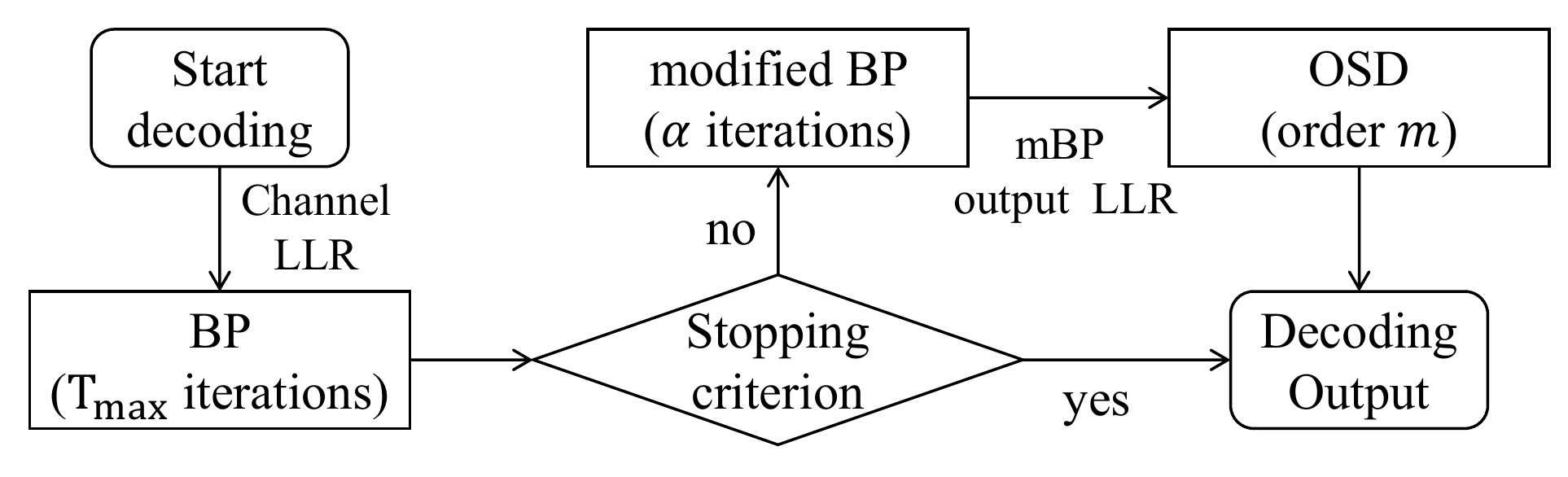}
\vspace{-0.5em}
\caption{The flowchart of the order-$m$ mBP-OSD algorithm}
\vspace{-1em}
\label{fig1}
\end{figure}

The overall procedure of the proposed order-$m$ mBP-OSD is demonstrated in Fig.~\ref{fig1}. At the beginning of mBP-OSD, the original BP decoding, as introduced in Section II-A, is performed over received channel LLR, i.e., $\bm{\ell}$. The output of BP, i.e., $\mathbf{x}$, is then verified by a designed stopping criterion, which determines if $\mathbf{x}$ can be claimed as the final decoding result. If not, mBP will be performed over $\bm{\ell}$ for a certain number (i.e., $\alpha$) of iterations, which fine-tunes the received sequence for the following OSD stage. Finally, an order-$m$ OSD is performed based on the output LLR of mBP, denoted by $\mathbf{L}'$, to obtain the near maximum-likelihood decoding result. 

In mBP-OSD, the stopping criterion can early terminate the decoding when BP is very likely to obtain the correct codeword estimate, which, therefore, can significantly reduce the overall decoding complexity. On the other hand, mBP is designed to reduce the error in MRPs that are input to OSD. This can either enhance error rate performance if the OSD order remains constant, or reduce the required OSD order if the error rate target is fixed.

The details of mBP-OSD will be introduced in the subsequent subsections.

\subsection{Stopping Criterion}
  
We proposed a stopping criterion for early terminating the mBP-OSD decoding after performing the original BP. Specifically, let $\mathbf{z}=[z_1,z_2...z_N]$ denote the hard decision of received message $\mathbf{r}$. Then, the WHD between the BP output $\mathbf{x}$ and $\mathbf{r}$ is defined as 
\begin{equation}
\mathcal{D}_{\mathbf{x}}=\sum_{\substack{0<i<N\\x_i\ne{z_i}}} |4r_i/N_0| = \sum_{\substack{0<i<N\\x_i\ne{z_i}}} |\ell_i|.
\end{equation}

The stopping criterion is described as follows. The BP output $\mathbf{x}$, is directly claimed as the decoding result, when 

\begin{equation}
\mathbf{x}\mathbf{H}^{\rm{T}} = 0  \ \ \ \text{and} \ \ \ \mathcal{D}_{\mathbf{x}}\leq \lambda,
\end{equation}
are both satisfied, where $\lambda$ is a given parameter.

The condition $\mathbf{x}\mathbf{H}^{\rm{T}} = 0$ is inspired by the optimality of BP for long LDPC codes. In \cite{b5} and \cite{b21}, when $\mathbf{x}\mathbf{H}^{\rm{T}} = 0$ is satisfied, the output of BP is regarded as the correct transmitted codeword. This is because for long LDPC codes, only a negligible proportion of BP decoding errors are undetected errors, i.e., errors still satisfying the code parity check. 

For short LDPC codes, even if $\mathbf{x}\mathbf{H}^{\rm{T}} = 0$ is satisfied (if $\mathbf{x}$ is a valid codeword), there is still a chance of BP decoding error, because short codewords have a higher probability of passing parity check due to the limited number of check nodes. Thus, we apply the condition $\mathcal{D}_{\mathrm{x}}\leq \lambda$ to further improve the accuracy of early termination.

We simulate the ratio between ratio of errors that (10) fails to identify to the total decoding errors for various half-rate LDPC codes at SNR = 1dB, which is summarized in Table \uppercase\expandafter{\romannumeral1}. As shown, a smaller $\lambda$ can achieve a higher early termination accuracy. We note that when $\lambda\to \infty $, (10) simplifies to only $\mathbf{x}\mathbf{H}^{\rm{T}}$ = 0. For very short code (e.g., $N = 32$), it is particularly critical to apply a small $\lambda$ to avoid the decoding performance degradation caused by false positive results of (10).

\begin{table}
\caption{The ratio of the number of errors satisfying equation (10) to total decoding errors at SNR = 1 dB}
\vspace{-1em}
\begin{center}
 \begin{tabular}{||c c c c||} 
 \hline
 LDPC codes & $\lambda=1$ & $\lambda=10$ & $\lambda\rightarrow+\infty$\\ [0.5ex] 
 \hline\hline
 $(32,16)$ & 1.47$\times 10^{-3}$ & 1.93$\times 10^{-2}$ & 1.99$\times 10^{-2}$ \\ 
 \hline
 $(96,48)$ & $< 10^{-5}$ & 8.23$\times 10^{-4}$ &1.83$\times 10^{-3}$ \\
 \hline
 $(128,64)$ &  $< 10^{-5}$ & $< 10^{-5}$ &3.84$\times 10^{-5}$ \\
 \hline
\end{tabular}
\end{center}
\vspace{-2em}
\end{table}

\subsection{Modified Belief Propagation Decoding}

When the stopping criterion described in Section \uppercase\expandafter{\romannumeral3}-B is not satisfied, mBP is performed based on the received LLR $\bm{\ell}$. 

The mBP applies an offset parameters $\beta$ to the extrinsic information passed from check nodes to variable nodes in each iteration. Specifically, the message passing of mBP is described as
\begin{equation}
M_{ji}^{'}=\ell_{i}+\beta(\sum_{j^{'}\in{\mathcal{A}_i}\backslash j}E_{ij^{'}}^{'}), 
\end{equation}
and
\begin{equation}
E_{ij}^{'}=2\tanh^{-1}(\prod_{i^{'}\in{\mathcal{B}_j}\backslash i}\tanh(M_{ji^{'}}^{'}/2)).
\end{equation}

Similarly, the posterior LLR after each mBP iteration is computed as
\begin{equation}
L_i^{'}=\ell_{i}+\beta(\sum_{j\in{\mathcal{A}_i}}E_{ij}^{'}).
\end{equation}

Different from the original BP conducted at the first stage of mBP-OSD, mBP only executes for $\alpha$ iterations, and then the posterior LLR $\mathbf{L}' = [L_1',\ldots,L_N']$ is output as the input of OSD. We note that the objective of mBP is to improve the quality of the OSD input, rather than to obtain a codeword estimate. This distinguishes it from the offset-BP approaches in \cite{b7,b8,b9,b10,b11} for improving the error rate of BP. In Section \uppercase\expandafter{\romannumeral4}-B, we will show that a small $\alpha$ suffices to significantly enhance the performance of mBP-OSD.

\subsection{Parameter Selection}

\subsubsection{The Selection of $\alpha$} 
The parameter $\alpha$ is the number of iterations of mBP. As described in Section \uppercase\expandafter{\romannumeral3}-C, the goal of mBP is to effectively reduce the number of errors in the MRPs of OSD input. Specifically, OSD can find the transmitted codeword only if all errors over MRPs are eliminated by a TEP \cite{b18}. Therefore, fewer errors in MRPs necessitate a TEP with a lower Hamming weight, resulting in a lower decoding order and reduced decoding complexity. The mBP iteratively updates the received LLR to improve its reliability by using extrinsic information. In this manner, the set of MRPs will be refined and thus the number of errors in MRPs can be reduced.

The value of $\alpha$ has a significant impact on the effectiveness of mBP. Precisely, mBP should incorporate as much extrinsic information as possible when refining the received LLR. On the other hand, due to the correlations introduced by graph cycles, excessive use of extrinsic information may also diminish the reliability of the refined LLR. Therefore, we proposed to determine $\alpha$ according to the girth $g$ of the code Tanner graph. Specifically, $\alpha$ is given by
\begin{equation}
\alpha= \lfloor{g/4+1}\rfloor.
\end{equation}
This selection of $\alpha$ ensures that mBP terminates at early iterations where correlations between passed messages barely occur. The girth $g$ can be determined theoretically for long LDPC codes \cite[Theorem 1]{b25}, or obtained numerically by enumeration for short codes. For example, one can apply the algorithm provided in \cite{b26} to obtain $g$, and accordingly the parameter $\alpha$ for a given code. Simulation results in Section \uppercase\expandafter{\romannumeral4}-B will show that the modified BP with $\alpha$ from (14) can effectively refine the MRPs to decrease the OSD order.

\subsubsection{The Selection of $\beta$}
\begin{figure}
\centering  

\subfigure[mBP-OSD of different decoding order at SNR = 3dB]{
\label{fig2.sub.1}
\includegraphics[width=1\linewidth]{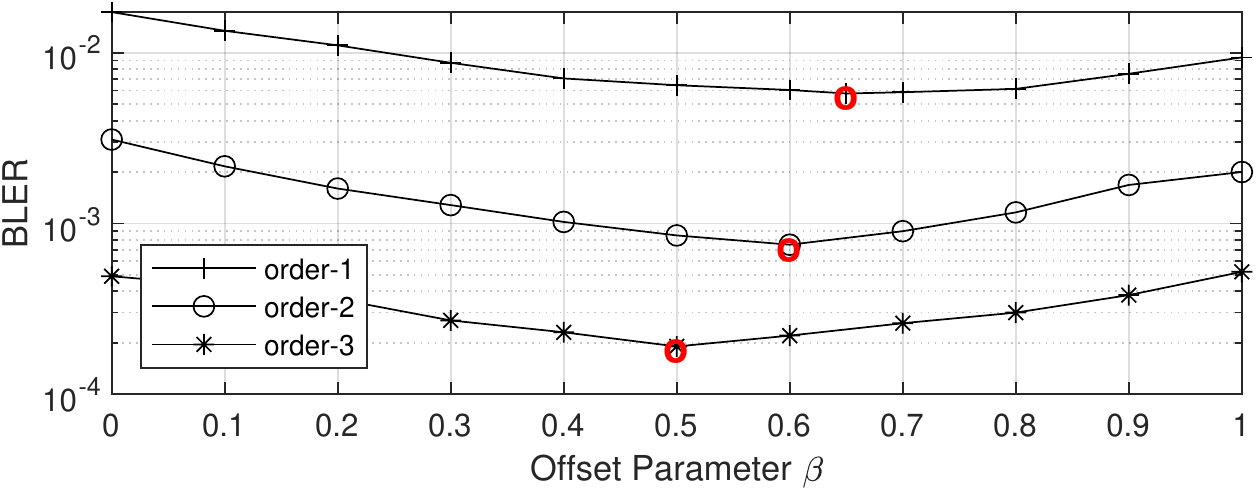}}
\subfigure[Order-1 mBP-OSD at different SNRs]{
\label{fig2.sub.2}
\includegraphics[width=1\linewidth]{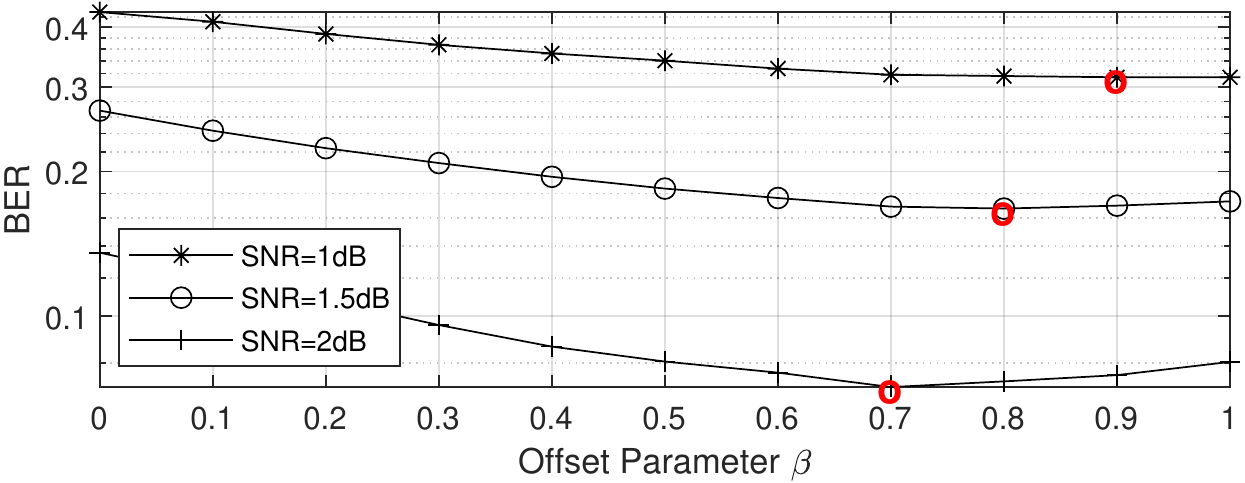}}
\caption{BLER of mBP-OSD for decoding (128,64) LDPC code with different $\beta$ values}
\vspace{-1em}

\label{fig2}
\end{figure}

During mBP iterations, the offset parameter $\beta$ adjusts the ratio between the internal and extrinsic information of variable nodes. In this paper, we experimentally find the optimal value of $\beta$ via simulations. Surprisingly, the optimal value of parameter $\beta$ is related to the decoding order of OSD. Fig.\ref{fig2.sub.1} illustrates the BLER performance of (128,64) CCSDS LDPC code with different $\beta$ values when $\lambda\to\infty$ and $\alpha=2$. As shown, for order-1, order-2, and order-3 mBP-OSD algorithms, the optimal $\beta$ values are 0.65, 0.6, and 0.5, respectively. It indicates that a larger OSD order requires a smaller $\beta$ to achieve the optimal BLER performance. 
Furthermore, Fig.~\ref{fig2.sub.2} shows that the optimal value of $\beta$ is also affected by the channel condition, i.e., it decreases as SNR increases. The optimal value of $\beta$ can be theoretically found by the density evolution analysis \cite{b5}, but a detailed exploration is beyond the scope of this paper.

\subsection{Algorithm of mBP-OSD}
The algorithm of mBP-OSD is presented in Algorithm \ref{alg:1}. The steps of BP are briefly stated in line 1 due to the space limit, and we refer readers to Section \uppercase\expandafter{\romannumeral2}-A for details.
    
\begin{algorithm}
\small    
\caption{order-$m$ mBP-OSD Decoder}  
\label{alg:1}  
\begin{algorithmic}[1]  
\REQUIRE Generator matrix $\mathbf{G}$, received sequence $\mathbf{r}$,  maximum number of iterations $\mathrm{T}_{\max}$, OSD decoding order $m$, parameters $\lambda$, $\alpha$ and $\beta$
\ENSURE Codeword estimation $\textbf{c}_{\mathrm{opt}}$
\\

\STATE Perform BP over the LLR  $\ell_i = 4r_i/N_0$, and output $\mathbf{x}$ 
\STATE Compute $\mathcal{D}_{\mathrm{x}}$ for $\mathbf{x}$ according to (9)
\IF {$\mathbf{x}\mathbf{H}^{\rm{T}} = 0$ and $\mathcal{D}_{\mathrm{x}}\leq \lambda$ are both satisfied}
\STATE Output $\mathbf{x}$ as $\mathbf{c}_{\mathrm{opt}}$
\ELSE

\STATE Initialize $E_{ij}'=0$ for $1\leq i \leq N$ and $1\leq j \leq N-K$.

\FOR{$t = 0:\alpha$}
\STATE Perform (11) for $1\leq i\leq N$ and $j\in{\mathcal{A}_i}$ to obtain $M_{ji}^{'}$
\STATE Perform (12) for $1\leq j \leq N\!-\! K$ and $i\in{\mathcal{B}_j}$ to obtain $E_{ij}^{'}$

\STATE Computing $L^{'} = [L_1....L_N]$ according to (13)

\ENDFOR

\STATE// OSD algorithm
\STATE First permutation $\pi_1$: $\ell^{'}=\pi_1(L^{'})$ and $\textbf{G}^{'}=\pi_1(\mathbf{G})$ 
\STATE Gaussian elimination and $\pi_2$: $\ell^{''}=\pi_2(\ell^{'})$ and $\mathbf{G}^{''}=\pi_2(\mathbf{G'})$ 
\STATE Perform hard-decision on $\ell^{''}$ to obtain $\mathbf{y}$

\FOR{$q=0:m$} 
\STATE Construct all TEPs with $q$ nonzero positions
\STATE Calculate $\mathbf{c}_{\mathbf{e}}=\mathbf{y}_\mathrm{e}\mathbf{G}^{''}=(\mathbf{y}_\mathrm{B}{\oplus}\mathbf{e})\mathbf{G}^{''}$ for each TEP $\mathbf{e}$

\ENDFOR
\STATE Obtain $\mathbf{c}_{\min}$ with the minimum WHD among all candidate codewords $\mathbf{c}_\mathrm{e}$
\RETURN $\mathbf{c}_{\mathrm{opt}}=\pi_1^{-1}(\pi_2^{-1}(\mathbf{c}_{\min}))$
\ENDIF
\end{algorithmic} 
\end{algorithm}

\section{Simulation Results and Analysis}
In this section, we analyze the decoding complexity and represent simulation results for short LDPC codes with different lengths. The maximum number of the original BP iterations is set to $\rm{T}_{\max} = 30$, and the number of the mBP iteration $\alpha$ is determined according to (14). We apply the optimal value of $\beta$ found via experiments and $\lambda$ which can achieve the ratio of undetected error to be less than $10^{-5}$.

\subsection{Complexity Analysis}

The decoding complexity of BP and OSD have been analyzed in \cite{b24} and \cite{b17}, respectively. According to \cite{b24}, a single BP iteration contains the following number of operations:
\begin{itemize}
\item Computing (1) or (3) requires an average of $N(N\!-\!K)$ additions, $8N(N\!-\!K\!-\!1)$ multiplications, and $N(N\!-\!K)$ divisions.
\item Computing (2) requires an average of $2N(N-K)$ additions and $3N(N-K-1)$ multiplications.
\end{itemize}
In general, the complexity of BP increases linearly with the number of edges of the Tanner graph, which is approximately represented as $O(N(N-K))$.

 For the order-$m$ OSD, the computational complexity is measured by numbers of floating point operations (FLOPs) and binary operations (BOPs) \cite{b17}. Specifically,
\begin{itemize}
\item The first permutation requires an average of $N\log(N)$ FLOPs \cite{b17}.
\item The Gaussian elimination requires an average of $N\min(K^2,(N-K)^2)$ BOPs \cite{b17}. 
\item The phase-$q$ reprocessing, $0\leq q\leq m$, of an order-$m$ OSD requires an average of $\tbinom{K}{p}(N-K)$ FLOPs and $\tbinom{K}{p}K(N-K)$ BOPs \cite{b17}. 

\end{itemize}
Consequently, an order-$m$ OSD has at least $\sum_{i=0}^{m}\binom{K}{i}(N-K)$ FLOPs and $\sum_{i=0}^{m}\binom{K}{i}K(N-K)$ BOPs. Thus, the complexity of OSD can be prohibitively high when the order $m$ is not small.

The proposed mBP-OSD can significantly reduce the complexity by skipping the OSD phase when the stopping criterion is satisfied. 
Let $\gamma$ denote the probability of the stopping criterion being satisfied by the BP output. The decoding complexity of an order-$m$ mBP-OSD, denoted by $\Gamma$, can be expressed as
\begin{equation}
\begin{aligned}
\Gamma\leq &(1-\gamma)\rm{T}_{\max}C_{\mathrm{BP}}+\gamma((\rm{T}_{max}+\alpha)C_{\mathrm{BP}}+C_{\mathrm{OSD}})\\
=& \rm{T}_{\max}C_{\mathrm{BP}}+ \gamma(\alpha C_{\mathrm{BP}}+C_{\mathrm{OSD}}),
\end{aligned}
\end{equation}
where $C_{\mathrm{BP}}$ and $C_{\mathrm{OSD}}$ are the decoding complexity of BP (similarly, mBP) per iteration and the order-$m$ OSD, respectively.

As shown by (15), when $\gamma$ is small, the complexity of mBP-OSD will be reduced to the level of BP, i.e., $\rm{T}_{\max}C_{\mathrm{BP}}$. The values of $\gamma$ for decoding the (96,48) TU KL LDPC code and (128,64) CCSDS LDPC code at different SNRs are recorded in Table \uppercase\expandafter{\romannumeral2}. It shows that $\gamma$ is very small at high SNRs, resulting in a low decoding complexity. We note that even at low SNRs ($\gamma$ is not small), mBP-OSD still has significantly lower complexity than OSD when achieving MLD performance, because $\rm{T}_{\max}C_{\mathrm{BP}} \ll C_{\mathrm{OSD}}$ for large OSD order $m$, and the order-$(m-1)$ mBP-OSD has the same BLER as the order-$m$ OSD. This will be further verified in Section \uppercase\expandafter{\romannumeral4}-B.

\begin{table}[t]
\caption{The value of the probability $\gamma$ at different SNRs}
\vspace{-1.5em}
\begin{center}
 \begin{tabular}{||c c c c c c||} 
 \hline
 SNR (dB) & 1 & 1.5 & 2 & 2.5 & 3 \\ [0.5ex] 
 \hline \hline
 $(128,64)$ LDPC code & 0.78 & 0.57 & 0.36 & 0.18 & 0.06\\ 
 \hline
 $(96,48)$ LDPC code & 0.60 & 0.40 & 0.24 & 0.11 & 0.04 \\
 \hline
\end{tabular}
\end{center}
\vspace{-2em}
\end{table}

\subsection{Simulation Results}

We include the original BP algorithm \cite{b2} and the original OSD algorithm \cite{b17} as the performance benchmarks in simulation. Meanwhile, we also compare with a decoder, termed BP-OSD, which simply concatenates BP and OSD. In BP-OSD, when BP decoding fails, the LLR output by BP will be directly fed into OSD for decoding. In addition, accumulated BP-OSD (aBP-OSD) proposed in \cite{b21} is also included for comparison. The maximum number of BP iterations is set to 30 in both BP-OSD and aBP-OSD. We compare the decoding complexity by comparing the average decoding time per codeword, which was obtained by averaging the total decoding time of 1000 codewords\footnote{In this paper, we implemented BP without parallelism. The decoding time of the BP component can be further reduced by applying parallelism.}. The simulation was conducted on Matlab 2022a with an Intel i7-12700H processor.

\begin{figure}
\centering  
\subfigure[BLER]{
\label{fig3.sub.1}
\includegraphics[width=0.47\linewidth]{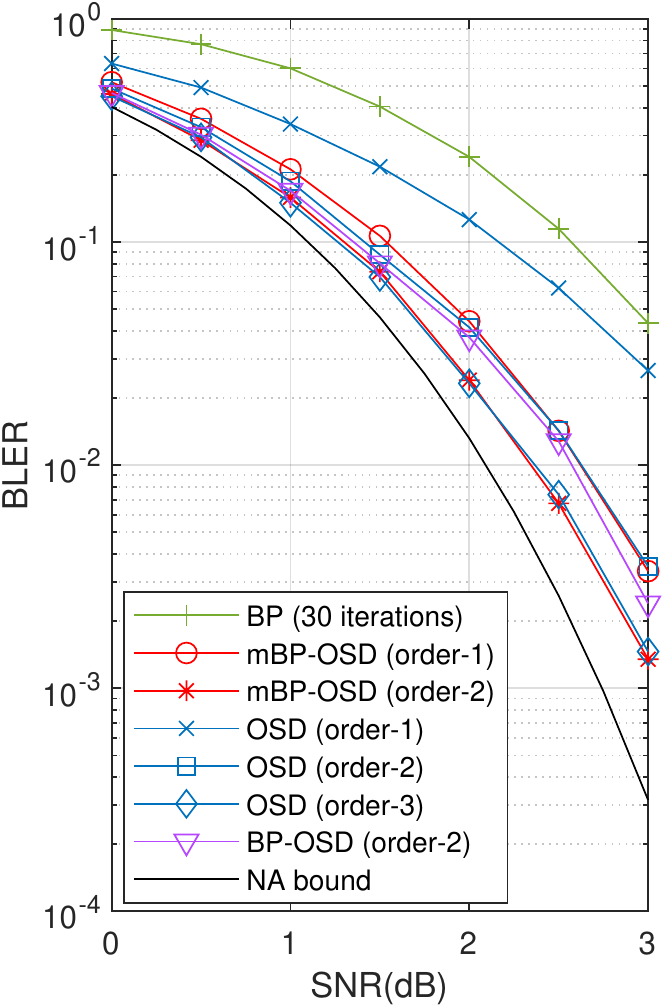}}
\subfigure[Complexity]{
\label{fig3.sub.2}
\includegraphics[width=0.47\linewidth]{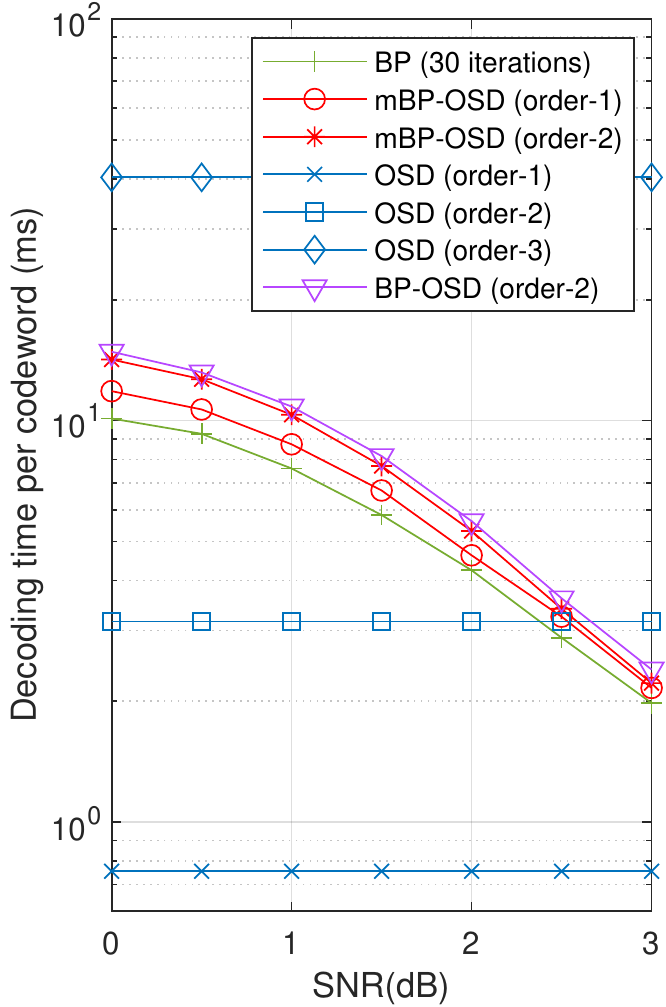}}

\vspace{-0.5em}
\caption{Performance and complexity comparison in decoding the (96,48) LDPC} 
\vspace{-1em}
\label{fig3}
\end{figure}

The performance and complexity for decoding the (96,48) TU KL LDPC code are depicted in Fig.~\ref{fig3}. As shown, the original BP has a significant BLER performance gap to the NA bound, while this gap is significantly reduced by the order-3 OSD (which is near MLD). The order-2 mBP-OSD achieves the same BLER as the order-3 OSD with a significantly reduced complexity. Specifically, as shown by \ref{fig3.sub.2}, the order-2 mBP-OSD has similar complexity to BP, which is 20 times lower than that of the order-3 OSD, and even lower than the order-2 OSD at high SNRs. BP-OSD is inferior to mBP-OSD in terms of both BLER and complexity, indicating the importance of mBP in the proposed algorithm.

\begin{figure}
\centering  
\subfigure[BLER]{
\label{fig4.sub.1}
\includegraphics[width=0.47\linewidth]{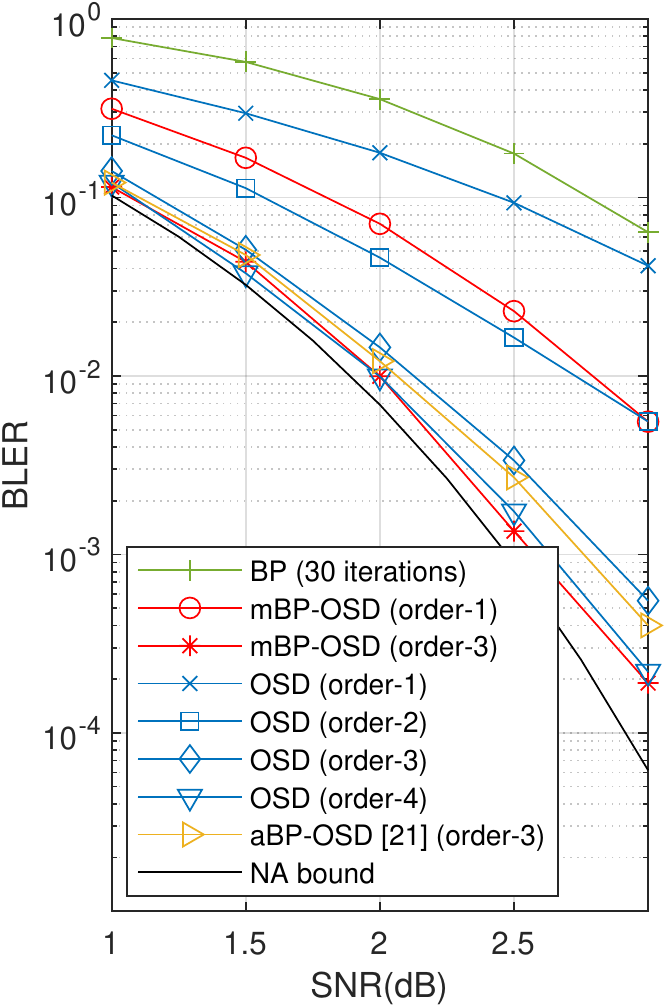}}
\subfigure[Complexity]{
\label{fig4.sub.2}
\includegraphics[width=0.47\linewidth]{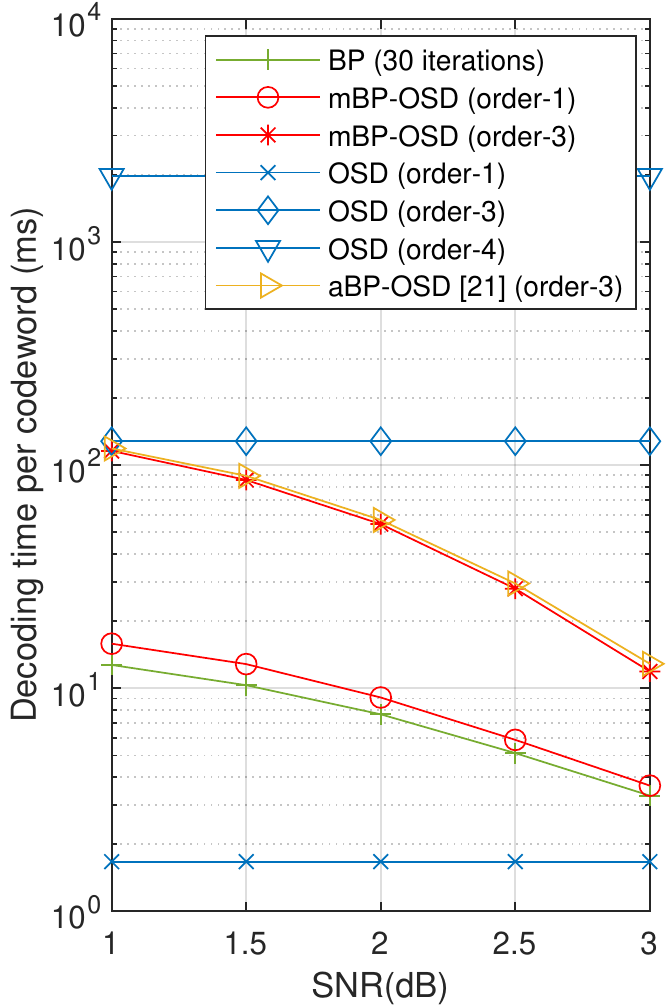}}

\vspace{-0.5em}
\caption{Performance and complexity comparison in decoding the (128,64) LDPC code}
\vspace{-1em}
\label{fig4}
\end{figure}

A similar performance gain can be observed for the (128, 64) CCSDS LDPC code, as illustrated in Fig.~\ref{fig4}. Precisely, the order-3 mBP-OSD can achieve the BLER performance of order-4 OSD (approximately MLD). In terms of complexity,  the order-3 mBP-OSD requires only 30 ms to decode one codeword, compared to 200 ms of the order-3 OSD and 3000 ms of the order-4 OSD at high SNRs. Compared with the aBP-OSD \cite{b21}, our proposed algorithm has a notable improvement in BLER performance while maintaining the same complexity.

\begin{figure}
\centering  
\subfigure[BLER]{
\label{fig5.sub.1}
\includegraphics[width=0.47\linewidth]{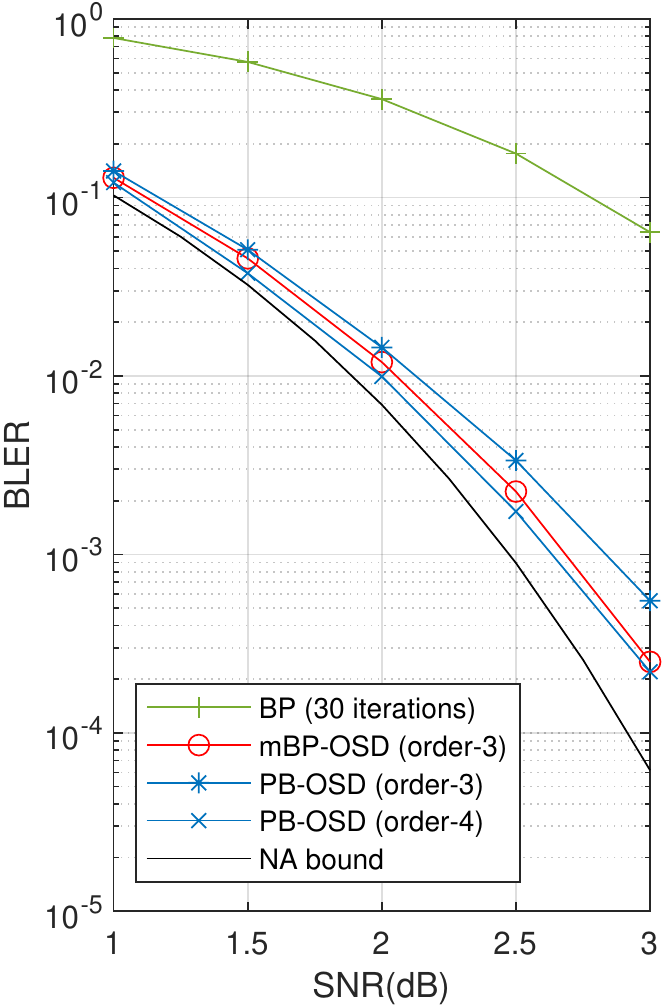}}
\subfigure[Complexity]{
\label{fig5.sub.2}
\includegraphics[width=0.47\linewidth]{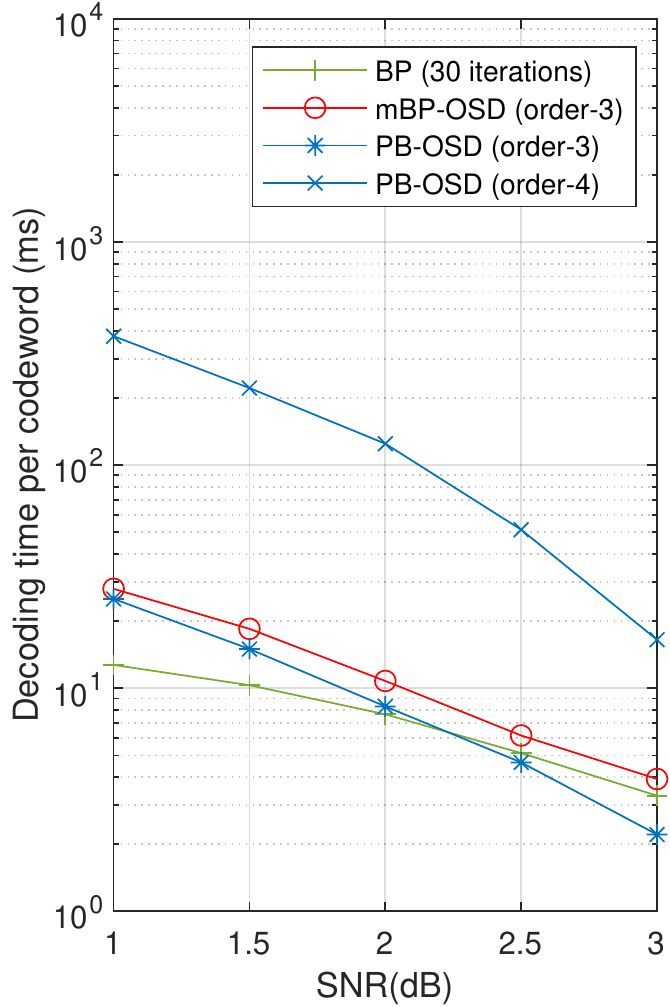}}

\vspace{-0.5em}
\caption{Performance and complexity comparison of the enhanced mBP-OSD (applying the PB-OSD from \cite{b18}) in decoding the (128,64) LDPC code} 
\vspace{-1em}
\label{fig5}
\end{figure}

We further replace the OSD stage in the proposed mBP-OSD by the latest OSD approach, probability-based OSD (PB-OSD) from \cite{b18}, to achieve an enhanced mBP-OSD\footnote{The optimal value of $\beta$ of the  enhanced mBP-OSD can be slightly different from that of the original mBP-OSD.}. PB-OSD has a significant lower complexity than OSD, while having the same BLER performance. With the (128, 64) CCSDS LDPC code, the BLER and decoding complexity of the enhanced mBP-OSD, BP, and the pure PB-OSD are compared in Fig.~\ref{fig5}. As shown, the order-3 mBP-OSD can still approximately achieve the BLER performance of the order-4 PB-OSD, which is close to the MLD performance. However, as shown by Fig.~\ref{fig5.sub.2}, the complexity of order-3 mBP-OSD is only slightly higher than BP and the order-3 PB-OSD, and significantly lower than that of the order-4 PB-OSD. These results indicate that the mBP-OSD can be combined with any improved OSD decoders to improve their BLER (at fixed complexity), or reduce their complexity (when approaching the MLD performance).

\vspace{-0.3em}
\section{Conclusion}
\vspace{-0.3em}
In this paper, we proposed a new modified-BP-OSD (mBP-OSD) algorithm for short LDPC codes, which carefully combines BP and OSD to approach the MLD performance. Two main techniques are applied: 1) A stopping criterion that can effectively early terminate mBP-OSD to significantly reduce its complexity and 2) a modified BP that effectively enhances the input reliability of OSD, thereby reducing the required OSD order or improving the BLER performance.

Simulation results show that the proposed mBP-OSD algorithm can approach MLD performance for short LDPC code with only a slight increase in complexity compared to BP at high SNRs. In addition, the order-$(m-1)$ mBP-OSD can achieve the same BLER as the order-$m$ OSD, demonstrating the capability to enhance any existing OSD-based approaches, such as PB-OSD \cite{b18}, for short LDPC codes in uRLLC.

\end{document}